\documentstyle[11pt,aaspp4]{article}
\begin{document}
\title{Signatures of the Origin of High-Energy Cosmic Rays in
Cosmological Gamma-Ray Bursts}
\author{Jordi Miralda-Escud\'e and Eli Waxman}
\affil{Institute for Advanced Study, Princeton, NJ 08540}
\authoremail{waxman@sns.ias.edu}
\slugcomment{Submitted to Ap. J. ({\it Letters})}

\begin{abstract}

  We derive observational consequences of the hypothesis that cosmic rays 
(CR's) of energy $>10^{19}{\rm eV}$ originate in the same cosmological 
objects producing gamma-ray bursts (GRB's). Inter-galactic magnetic fields 
$\gtrsim 10^{-12} {\rm G}$ are required in this model to allow CR's to be 
observed continuously in time by producing energy dependent delays in the 
CR arrival times. This results in individual CR sources having very narrow
observed spectra, since at any given time only those CR's having a fixed
time delay are observed. Thus, the brightest CR sources should be different 
at different energies. The average number of sources contributing to the
total CR flux decreases with energy much more rapidly than in a model of 
steady CR sources, dropping to one at $E_{\rm crit} \simeq2\times10^{20}$~eV 
with very weak sensitivity to the inter-galactic magnetic field strength. 
Below $E_{\rm crit}$, a very large number of sources is expected, consistent 
with observations. Above $E_{\rm crit}$, a source may be observed with a flux 
considerably higher than the time-averaged CR flux from all sources, if a 
nearby GRB occurred recently. If such a source is present, its narrow spectrum 
may produce a ``gap'' in the total spectrum. These signatures should be 
detectable by the planned ``Auger'' CR experiment.

\end{abstract}

\section{Introduction}

The sources of gamma ray bursts (GRB's) and of cosmic rays (CR's) with energy
$E>10^{19}{\rm eV}$ are unknown. In particular, most of the sources
of cosmic rays that have been proposed have difficulties in accelerating
CR's up to the highest observed energies (\cite{Hillas,huge}).
Recent gamma ray and cosmic ray observations give increasing evidence
that both phenomena are of cosmological origin [see \cite{cos1}, \cite{cos2}, 
\cite{cos3} for GRB's; \cite{Fly1}, \cite{AGASA2}, \cite{Wb} for CR's].
Although the source of GRB's is unknown, their observational
characteristics impose strong constraints on the physical conditions
in the $\gamma$-ray emitting region (\cite{scen1}, \cite{scen2}),
which imply that protons may be accelerated in the $\gamma$-ray emitting 
region to energies $10^{20} - 10^{21} {\rm eV}$ (\cite{Wa}, \cite{Vietri}).
In addition, the average rate (over volume and time) at which energy
is emitted as $\gamma$-rays by GRB's and in CR's above $10^{19}
{\rm eV}$ in the cosmological scenario is, remarkably, comparable 
(\cite{Wa},b). These two facts suggest that GRB's and high-energy CR's may 
have a common origin.

  An essential ingredient of a bursting model for cosmic rays is the
time delay due to inter-galactic magnetic fields. The energy of the most
energetic CR detected by the Fly's Eye experiment is in excess of
$2\times10^{20}{\rm eV}$ (\cite{Fly}), and that of the most
energetic AGASA event is above $10^{20}{\rm eV}$ (\cite{AGASA1}). On a
cosmological scale, the distance traveled by such energetic particles is
small: $<100{\rm Mpc}$ for the AGASA event, $<50{\rm Mpc}$ for the Fly's Eye
event (\cite{dist}). Thus, the detection of these events over a $\sim5
{\rm yr}$ period can be reconciled with the rate of nearby GRB's ($\sim1$
per $50{\rm yr}$ in the field of view of the CR experiments out to $100
{\rm Mpc}$ in a standard cosmological scenario; e.g., \cite{rate2}) only if
there is a large dispersion in the arrival time of protons produced in a
single burst. The required dispersion, $\geq50
{\rm yr}$ for $10^{20}{\rm eV}$ proton, may be produced by inter-galactic
magnetic fields (\cite{Wa,Coppi}).
The deflection angle for a proton
propagating a distance $D$ in a magnetic field $B$ with
coherence length $\lambda$ is $\theta_s \simeq 0.05\arcdeg\,
(D/\lambda)^{1/2}\, (\lambda/10{\rm Mpc})\, (B/10^{-11}\, {\rm G})\,
(E/10^{20}{\rm eV})^{-1} $, and the induced 
time delay is $\tau(E) \simeq 10^3 {\rm yr}
(D/100{\rm Mpc})^2\, (\lambda/10{\rm Mpc})\, (B/10^{-11}\, {\rm G})^2\,
(E/10^{20}{\rm eV})^{-2} $. Since the time delay is
energy dependent, the large spread in proton energy, induced by random
energy loss, results in a time broadening of the CR
pulse over a time $\sim\tau(E)$. Thus, the CR's from a single
burst can be received on Earth over a long time interval. Nevertheless,
since the angular deflection is small,
the individual sources are still detectable by measuring the arrival
directions.

  In this {\it Letter}, we examine the characteristics of CR sources that 
should be expected in a bursting source model, as a
function of the time delay, dependent on the magnetic field.
We find that there are characteristic signatures for such a model,
which would allow to distinguish it from a scenario where the CR
sources are steady, i.e., where the sources emit a constant flux on a
time scale longer than the time delay of the lowest energy CR's that
are relevant. In \S 2.1 we give a qualitative description of the 
bursting model properties, using an approximate analytic approach.
In \S 2.2 we present Monte-Carlo simulations that demonstrate the
properties discussed in \S 2.1.
The various tests of the bursting model and
the implications for future high energy CR experiments are discussed in
\S 3. 

\section{Characteristics of Cosmic Ray Bursts Sources}

  We consider a Cosmic Ray Burst (CRB) taking place at a distance $D$
from us, where a total number of protons $n_p(E){\rm d}E$ of
energy $E$ is emitted at a single instant in time. The CR's
arrive with a time delay $t$, relative to gamma-rays, with a
probability density $p[t/\tau(D,E)]{\rm d}[t/\tau(D,E)]$, where
$\tau(D,E) \propto D^2/E^2$ is the characteristic time delay. The time
delay $t$ at a fixed energy and distance varies randomly due to two
effects. First, the magnetic field along a trajectory should have random
variations; for example, if the magnetic fields originate in galaxies
and are later ejected to the inter-galactic medium, the field strength
along a trajectory should vary depending on the impact parameter to
individual galaxies. In the absence of energy losses, the bursting
source would produce a number of cosmic ray ``images'', and the cosmic
rays in each image would be of a single energy which would decrease with
time as $t^{-1/2}$. However, the random nature of the energy loss of a
cosmic ray eliminates these images, and simply introduces a dispersion
in the arrival times and arrival directions at a fixed energy. In
general, the dispersion in arrival times $t$ will be of order
$\tau(D,E)$ (\cite{Coppi}).

\subsection{Analytic Model}

  We now perform a simple analytic calculation of the number of CR
sources that should be seen at each energy and flux in a CRB model.
For this purpose,
we approximate the effect of energy losses as being negligible
when CR's come from a distance $D < D_c(E)$, and eliminating all
cosmic rays coming from $D > D_c(E)$
(for $E<10^{20}{\rm eV}$, this approximation is quite good and
$D_c(E)$ corresponds to the distance where the initial proton energy
necessary to have an observed energy $E$, after losses to
electron-positron production, exceeds the threshold for pion production).
We also assume that the sources
are observed only during a time $\tau(D,E)$ with a constant flux
\begin{equation}
F(E,D) = {n_p(E) \over 4\pi D^2 \tau(D,E)} = 
{n_p(E) D_c(E)^2 \over 4\pi D^4 \tau_c(E)}\quad,
\label{flux}
\end{equation}
where $\tau_c(E) = \tau(D_c(E),E)$.
If the rate per unit volume
of CRB's is $\nu$,
all emitting the same $n_p(E)$, then the average number of bursts
at distance $D$ observed at any time is $n(D,E)\, {\rm d}D = 4\pi\nu
\tau_c(E) [D^4/D_c(E)^2]\, {\rm d}D$, giving a number of bursts at a given
observed flux
\begin{equation}
n(F,E)\, {\rm d}F = \pi\nu D_c(E)^3\tau_c(E) \left[ F_c(E)\over F
\right]^{5/4}\, {{\rm d}F\over F}\quad, 
\label{n}
\end{equation}
where $F_c(E) = n_p(E)/[4\pi D_c(E)^2 \tau_c(E)]$. The flux $F_c(E)$ is
the minimum flux observed for the sources.
In our simplified model, the number of sources drops to zero abruptly
at $F_c(E)$ owing to the assumed distance cutoff $D_c(E)$ and the
``top-hat'' time profile. In reality,
there should be a smooth turnover at $F_c(E)$ of the number of CRB
sources from the $-5/4$ power-law slope at the bright end.
This result for bursting sources is in contrast to the usual $-3/2$
Euclidean slope, which applies for steady sources of cosmic rays.

  The total average number of sources above flux $F$ is
\begin{equation}
N(F,E) = {4\pi\nu\over 5} D_c(E)^3 \tau_c(E)\left[
{F_c(E)\over F}\right]^{5/4}\quad,
\label{N}
\end{equation}
and the average background flux resulting from all the sources is
\begin{equation}
B(E) = 4\pi\nu D_c(E)^3\tau_c(E)F_c(E) = \nu n_p(E) D_c(E)
\quad.
\label{B}
\end{equation}
The background flux
is dominated by sources with flux near $F_c(E)$, although the
contribution from brighter sources decreases only as $F^{-1/4}$.

  As the cosmic ray energy is increased, the average number of bursts
observed above the turnover flux $F_c(E)$ decreases, and there is a
critical energy $E_{\rm crit}$ where this average number of sources equals
unity:
\begin{equation}
{4\pi\nu \over 5}D_c(E_{\rm crit})^3 \tau_c(E_{\rm crit}) = 1
\quad.
\label{Ec}
\end{equation}
We can write the average number of sources in terms of $E_{\rm crit}$ as
\begin{equation}
N_c(E) \equiv N[F_c(E), E] = \left( E_{\rm crit}\over E \right)^2
\left[ D_c(E)\over D_c(E_{\rm crit}) \right]^5 \quad.
\label{Nc}
\end{equation}

  The number of sources $N_c$ drops rapidly with energy, due to the strong
dependence on the decreasing cutoff distance $D_c(E)$.
The drop is especially rapid near $10^{20}{\rm eV}$, where $D_c(E)$
decreases quickly (see Fig. 2 below). Therefore, for $E < E_{\rm crit}$,
the number of sources contributing to the flux is very large, 
and the total number of CR's received at any
given time is near the average background $B(E)$. The brightest
source has a typical flux $F_1(E)\sim F_c(E)N_c(E)^{4/5}=
[B(E)/5](E/E_{\rm crit})^{2/5}[D_c(E)/D_c(E_{\rm crit})]^{-1}$, although
there is a probability to observe a source with $F > F_1(E)$,
$P \sim [F/F_1(E)]^{-5/4}$. At $E > E_{\rm crit}$, the total energy received
in CR's will generally be much lower than the average $B(E)$,
because there will be no burst within a distance $D_c(E)$ having taken
place sufficiently recently. The few CR's may be the lucky
survivors from sources further than $D_c(E)$, or they may have
anomalously long time-delays as a result of crossing a region of high
magnetic field (probably near a galaxy). There is, however, a
probability $P \simeq N_c(E)$ of seeing one CR source with
$E > E_{\rm crit}$ having a flux $\sim B(E)/N_c(E)$, or an even brighter
one with probability decreasing as $F^{-5/4}$.

  If the CR sources are steady, then the number of sources decreases with 
energy only as $D_c(E)^3$, i.e., much more slowly than predicted by eq. 
(\ref{Nc}). This implies that for a given critical energy, the number of 
bright sources at $E<E_{\rm crit}$ predicted by a model of steady
sources is much larger than that predicted for bursting sources. 

  Bursting CR sources should have narrowly peaked energy
spectra, and therefore the brightest sources should be different at
different energies. For example, if a bright source is observed at
$E > E_{\rm crit}$, the burst must have taken place recently in order that
the high energy cosmic rays are just arriving on Earth, so lower energy
cosmic rays will not have arrived yet. Typically, there will be other
brighter sources at $E < E_{\rm crit}$, corresponding to
bursts that took place a longer time ago, and probably closer to us
(since there is a longer time interval available). This is in marked
contrast to a model of steady state sources, where the brightest
source at high energies should also be the brightest one at low
energies, its fractional contribution to the background decreasing to
low energy only as $D_c(E)^{-1}$.

\subsection{Numerical Results}

  We now present the results of a Monte-Carlo simulation of the total
number of cosmic rays received from CRB's at some fixed time. For each
realization we randomly draw the positions (distances from Earth) and
times at which cosmological CRB's occured, 
assuming that the CRB's are homogeneously distributed
standard candles with an average rate $\nu = 2.3\times 10^{-8} h^{3}
{\rm Mpc}^{-3} {\rm yr}^{-1}$ (with $h=0.75$) similar to
that fitted to the observed flux distribution of GRB's assuming a
no-evolution standard candle model (\cite{rate2}).
We assume an intrinsic
cosmic ray generation spectrum $n_p(E) \propto E^{-2}{\rm d}E$, which
produces a flux above $2\times10^{19}{\rm eV}$ consistent with the
Fly's Eye and AGASA data (Waxman 1995b). We calculate 
the change of the spectrum due to interaction
with the CMB photons in a method similar to that described in Waxman (1995b),
except that for distances $<130{\rm Mpc}$ 
we do not use the continuous energy loss
approximation but rather an exact calculation of the energy loss, 
which includes fluctuations.

  For the probability distribution of the time-delay
for a cosmic ray of fixed energy from a source at a given distance, we
use the form $p(\tau) = p_0(\tau/\tau_0)^{-\alpha-1}$ for
$\tau > \tau_0$, $p(\tau){\rm d}\tau = (4p_0/3) (\tau/\tau_0 - 1/4)$ for
$\tau_0 > \tau > \tau_0/4$, and $p(\tau)=0$ for $\tau < \tau_0/4$, where
$\tau_0$ is the characteristic time-delay and is proportional to
$D^2/E^2$, and $p_0 = [ \tau_0(3/8 + 1/\alpha)]^{-1}$.
For $\tau < \tau_0$ the form of the probability function
approximately matches results obtained by Waxman \& Coppi (1995).
Cosmic rays with large $\tau$
are the ones that have crossed regions of high magnetic field, and we
assume there is a power-law distribution of magnetic field strengths,
giving a power-law distribution of time delays. If the typical field in
the inter-galactic medium is $B$ and has coherence length $\lambda$, the
typical deflection angle is $\theta \propto B\, (D\lambda)^{1/2}$. When
intercepting a region with magnetic field $B' \gg B$ and coherence
length $\lambda'$, the deflection angle is $\theta' \propto B' \lambda'$,
yielding a time delay $\tau' \propto \tau (B' \lambda'/ B \lambda)^2
(\lambda / D)$. If $n$ is the number density of such regions, the
interception probability is $n \pi \lambda'^2\, D$, so the index
$\alpha$ is $\alpha = - \log(\pi n \lambda'^2 D) / \log[(B'^2 \lambda'^2) /
(B^2 \lambda D) ]$. Here, we shall use as an example 
$\tau_1\equiv\tau(D=80{\rm Mpc}, E=10^{20}{\rm eV})=10^3$ years, 
corresponding
to $\lambda\approx 10{\rm Mpc}$ and $B\approx10^{-11} {\rm G}$.
We also take
$B'/B = 10^5$, $\lambda'/\lambda=10^{-3}$, and $n=10^{-2} {\rm Mpc}^{-3}$,
giving typical parameters for spiral galaxies (this leads to
$\alpha \simeq 1$, with a weak dependence on distance).

  We have examined a total of 50 realizations. About 70\% of these are
similar to the realization presented in Fig. 1a: there is no
source sufficiently nearby having occurred sufficiently recently, so
the flux at high energies is below the average. In the other $30\%$,
there are typically one or two bright sources dominating at
$E > 10^{20} {\rm eV}$, as in the realization presented in Fig. 1b. 
A source similar to the brightest one in Fig. 1b
appears only $4\%$ of the time (in this example, the source is at
$z=0.0056$ and occurred 51 years ago);
the second brightest source at $E\simeq 10^{20}$ eV
in Fig. 1b is more common. The analytic expression (\ref{N}) for
$N(F,E)$ provides a good approximation to the numerical results for
the number of sources with flux $F$ and spectral peak at $E$, except
that sources at high energy are also present at fluxes below $F_c(E)$,
coming from CRB's at distances higher than $D_c(E)$ for which some high
energy cosmic rays still survive. The spectral shape of the individual
sources is determined by the time-delay probability distribution
we have assumed, and is slightly modified by the interaction with the
microwave background (this is the reason why the shape of the spectra
of different sources varies).

  Fig. 2 shows $N_c(E)$, calculated from the average background and
equations (\ref{B}) and (\ref{Nc}). Since our numerical model does not assume
a sharp cutoff in the flux distribution, as we did above in the
analytical model, $N_c(E)$ is here an indication of the number of
sources above the turnover flux, which dominate the contribution to the
average background. 
All the characteristics of the sources depend on the CRB rate $\nu$ and
on the characteristic time delay
$\tau_1\equiv\tau(D=80{\rm Mpc}, E=10^{20}{\rm eV})$ only through their
product $\nu\tau_1$, or, equivalently, 
through the critical energy $E_{\rm crit}(
\nu\tau_1)$. For the parameters we have chosen, $\nu\tau_1=10^{-5}{\rm Mpc}
^{-3}$, Fig. 2 shows that $E_{\rm crit} \simeq
1.4\times 10^{20}$ eV. The dependence of $E_{\rm crit}$ on $\nu\tau_1$
is easily determined from Fig. 2, since $N_c\propto\nu\tau_1$ 
(see eqs. \ref{Ec}-\ref{Nc}) and therefore the curve in Fig. 2
shifts vertically as $\nu\tau_1$. If $>10^{19}{\rm eV}$
CR's are indeed produced by GRB's,
then $\nu$ is determined by the GRB flux distribution. The time delay,
however, depends on the unknown properties of the inter-galactic magnetic
field, $\tau_1\propto B^2 \lambda$. As mentioned in \S 1, current data requires
$\tau_1\gtrsim 50{\rm yr}$, or, equivalently, $E_{\rm crit} \gtrsim 
10^{20}$ eV, which corresponds to $B\lambda^{1/2}\gtrsim10^{-11}{\rm G\ Mpc}
^{1/2}$. The current upper limit for the inter-galactic magnetic field,
$B\lambda^{1/2}\le10^{-9}{\rm G\ Mpc}^{1/2}$ (\cite{Kron}, \cite{Vallee}),
allows a much larger delay, $\tau_1\le10^6{\rm yr}$. However, the rapid
decrease of $N_c(E)$ with energy near $10^{20}{\rm eV}$, implies that
$E_{\rm crit}$ is not very sensitive to $\nu\tau_1$. 
Thus, for the range allowed
for the GRB model, $5\times10^{-7}{\rm Mpc}^{-3}\le\nu
\tau_1\le10^{-2}{\rm Mpc}^{-3}$,
the critical energy is limited to the range $10^{20}{\rm eV}\le
E_{\rm crit}\le3\times10^{20}{\rm eV}$.

\section{Discussion}

   We have analyzed a model where $>10^{19}{\rm eV}$ CR's are produced 
by cosmological sources bursting at a rate comparable to GRB's. We have
found that, in this model, the average number of CR sources 
contributing to the flux decreases with energy much more rapidly than in the 
case where the CR sources are steady. We have shown that a critical energy 
exists, $10^{20}{\rm eV}\le E_{\rm crit}<3\times10^{20}{\rm eV}$,
above which a few sources produce most of the CR's, and that the 
observed spectra of these sources is very narrow: the bright sources
at high energy should be totally absent in cosmic rays of substantially
lower energy, since particles take longer to arrive the lower their
energy. In contrast, a model of steady sources predicts that the
brightest sources at high energies should also be the brightest ones at
low energies.

  Above $E_{\rm crit}$, there is a significant probability to observe one 
source with a flux considerably higher than average. If such a source is 
present, its narrow spectrum may produce a ``gap'' in the overall 
spectrum, as in Fig. 1b. 
Recently, \cite{TD} argued that the observation of such an energy gap would 
imply that the sources of $>10^{20}{\rm eV}$ CR's are different from
the sources at lower energy, hinting that the highest energy CR's are 
produced by the decay of a new type of massive
particles. We see here that this is not the case when bursting sources
are allowed, owing to the time variability. If such an energy gap is
present, our model predicts that most of the cosmic rays above the gap
should normally come from one source.

   If our model is correct, then the Fly's Eye event above 
$2\times10^{20}{\rm eV}$ suggests that we live at one of the times 
when a bright source is present at high energies. However, the absence of
such a source can not be ruled out, since, for example, the probability 
to have detected the Fly's Eye event in the realization of Fig. 1a, 
where no bright source exists, is $\sim 3\%$.
The highest energy AGASA event might more easily be produced
by a common, faint source (like in Fig. 1a). Furthermore,
notice that, given that Fly's Eye has
detected only one cosmic ray with $E > 10^{20}$ eV, we already know that
the AGASA cosmic ray had a low probability of being detected; within
the measurement error, its
energy might be not much above $10^{20}$~eV.

  Given the present scarcity of ultra-high energy CR's, no solid
conclusions can be drawn. However, with the projected Auger experiment 
(\cite{huge}), the number of detected CR's would increase by
a factor $\sim 50$. If $E_{\rm crit}$ is $\sim2\times10^{20}{\rm eV}$,
as predicted by our model, then a few bright sources 
above $10^{20}{\rm eV}$ should be identified. In addition, for $E_{\rm crit}=2
\times10^{20}{\rm eV}$ our model predicts $\sim10$ sources producing
more then $5$ events at $5\times10^{19}{\rm eV}$, compared to $\sim100$
such sources predicted in a steady source model with a similar $E_{crit}$.

  The observed characteristics of high energy 
CR sources depend on the bursting rate $\nu$ and on the typical time delay 
$\tau_1$ only through their product $\nu\tau_1$. However, $\nu$ and $\tau_1$ 
may be measured separately, if the time delay is either very short,
$\tau_1\leq50{\rm yr}$, or very long, $\tau_1\sim10^6{\rm yr}$. In the former
case, time variability of high energy sources may be detected, while
in the latter, which implies large magnetic fields, dispersion
in CR arrival directions could be measured. The magnetic field of our
galaxy can also have interesting observable effects: the images of CRB
sources should appear elongated perpendicular to the direction of the
magnetic field, with a predictable correlation of the cosmic ray
position and energy. For example, a cosmic ray with $E = 3 \times 10^{19}
{\rm eV}$ could be deflected by $\sim 10\arcdeg$ when arriving along the plane
of the galaxy.

  The positions of cosmic rays could also be correlated with
those of nearby galaxies to see if the events producing them occur in
normal stellar populations (\cite{Fisher}). 
The identification with GRB's could then
lead to further constraints on the nature of the objects producing
these explosions.

\acknowledgements
We thank K. Fisher for helpful discussions. This research
was partially supported by a W. M. Keck Foundation grant 
and NSF grant PHY92-45317 to the IAS, and by NSF grant PHY94-07194
to the ITP (UC Santa-Barbara).

\newpage
\begin{figure}
\plottwo{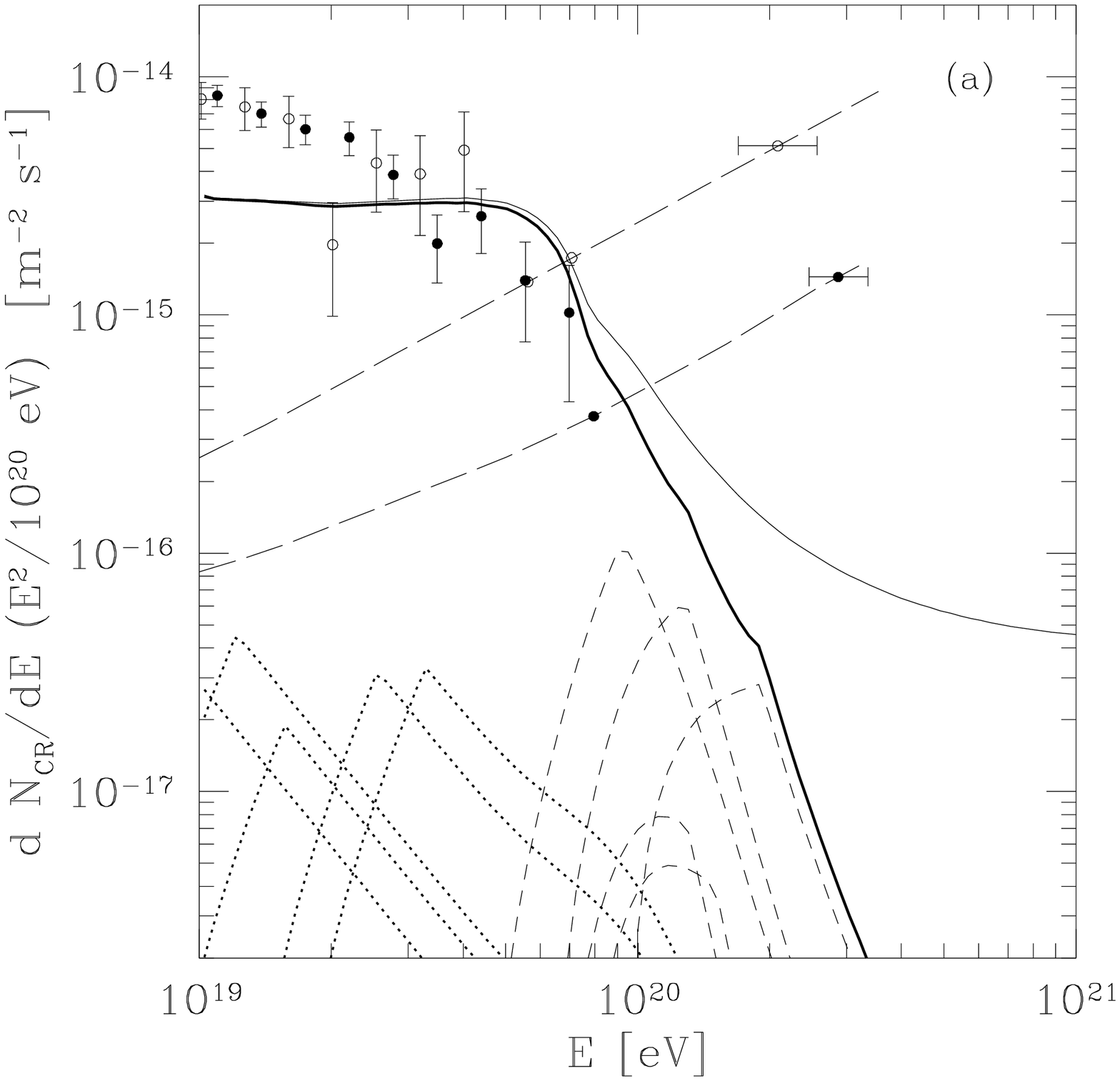}{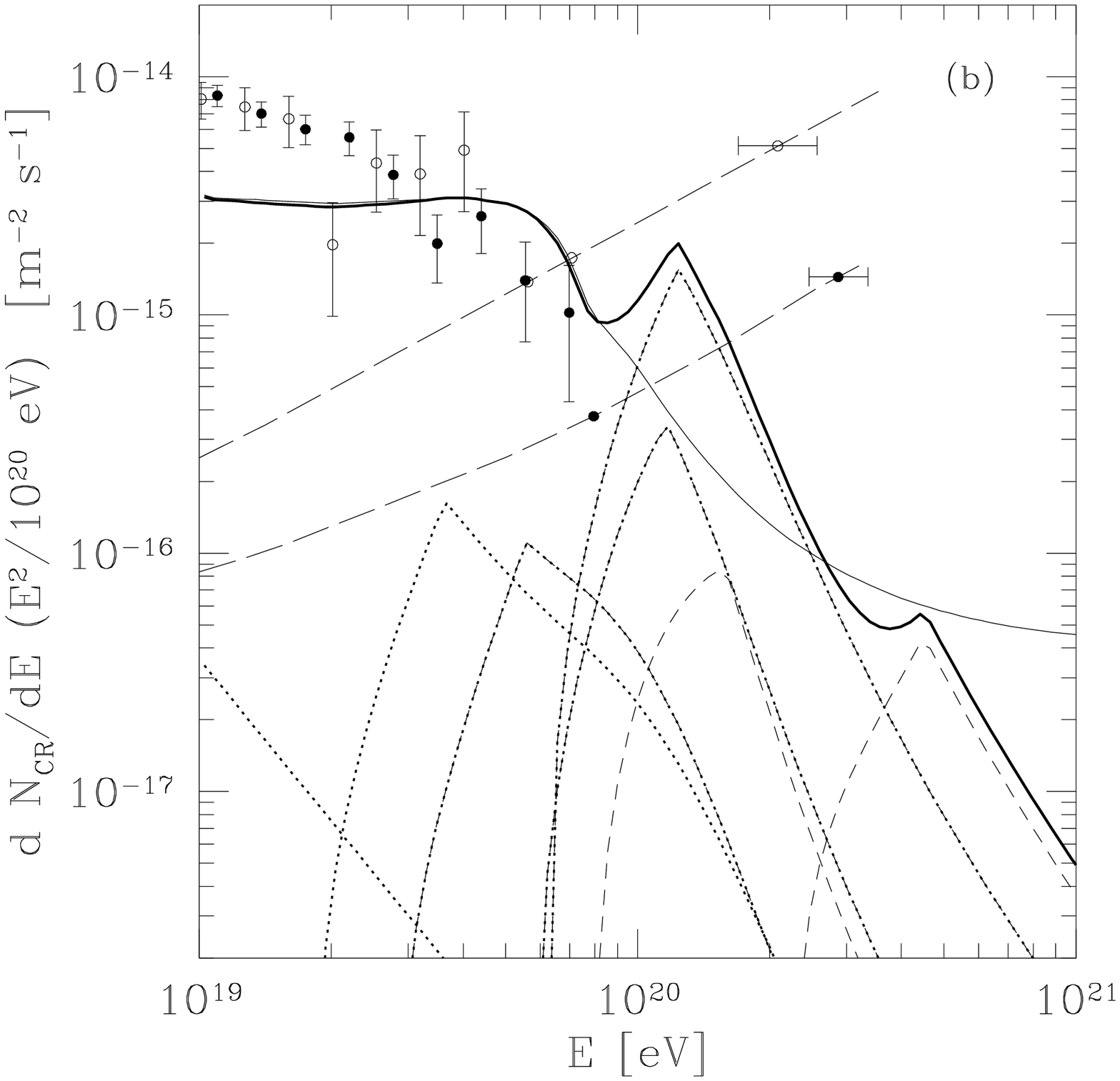}
\caption{Results of two Monte-Carlo realizations of the bursting sources
model with $\nu\tau_1=10^{-5}{\rm Mpc}^{-3}$: Thick solid line- overall 
spectrum in the realization,
shown as the number of cosmic rays per unit $\log E$, times
the energy. Thin solid line- average spectrum, obtained when the emissivity
is spatially uniform and not due to discrete sources; notice that this
curve is also proportional to $D_c(E)$, from eq. (4).
Dotted lines- spectra of the five sources having the largest CR flux.
Short dashed lines- spectra of the five sources that reach the highest
fraction of the average  flux.
Filled circles- Fly's Eye data. Open circles- AGASA data ($1\sigma$
errors are shown for the flux in bins with more than $1$ detected events,
and for the energy of the highest energy CR's).
Long dashed lines- the intensity where each experiment should have
detected on average one CR in each bin, where the bins are
equally spaced in $\log E$ (the upper line corresponds to
AGASA). The normalization of the average flux is chosen to
fit the observations at $E>2\times 10^{19}$ eV [at lower energies,
a contribution from iron cosmic rays from other sources is likely to
be present (Bird {\it et al.} 1994, Waxman 1995b)].}
\label{fig1}
\end{figure}

\begin{figure}
\plotone{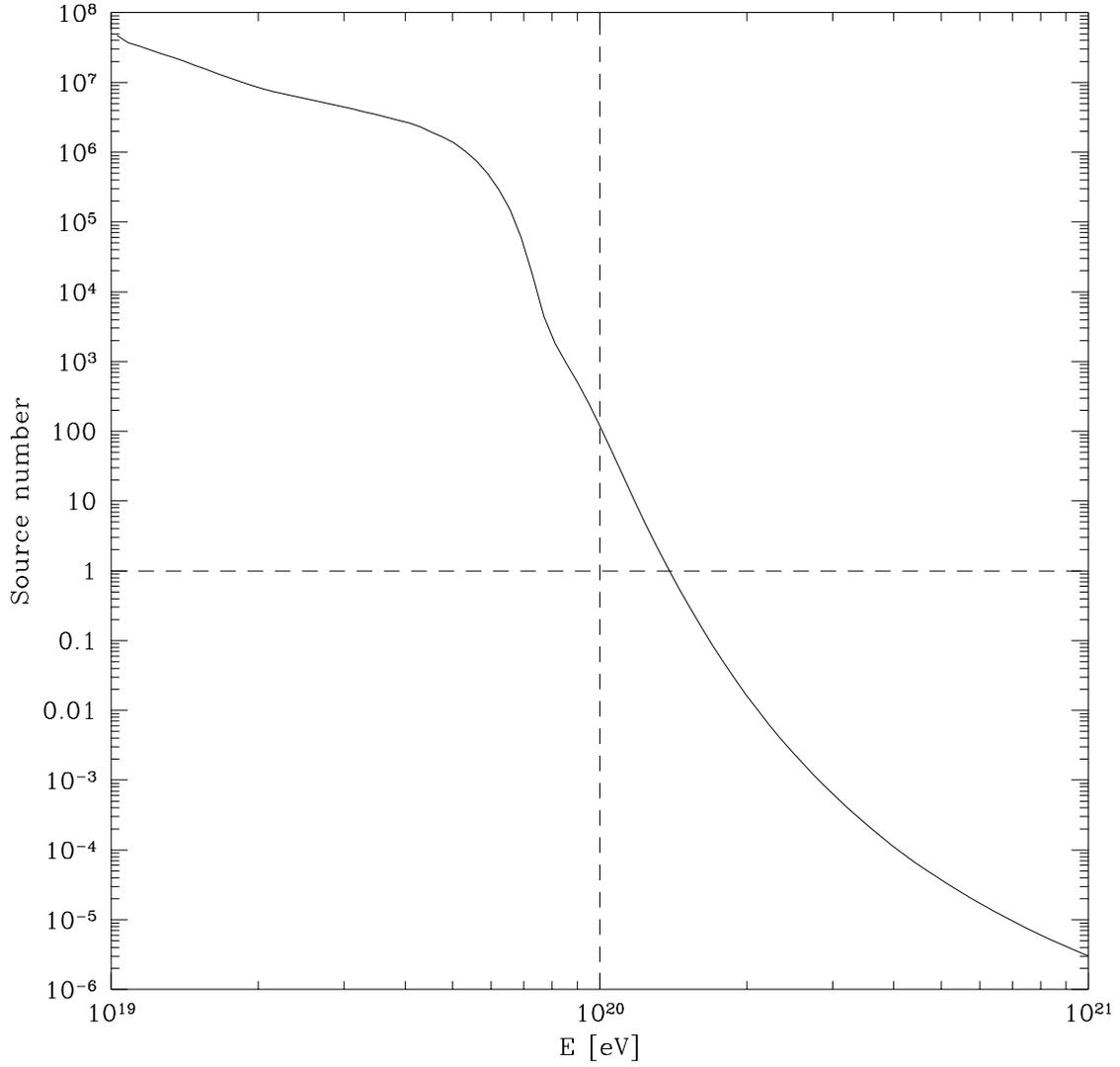}
\caption{
The average number of CR sources as function of energy, for bursting sources
model with $\nu\tau_1=10^{-5}{\rm Mpc}^{-3}$ 
(the dashed lines are added only for visual aid).}
\label{fig2}
\end{figure}

\end{document}